\documentclass[12pt,tightenlines,eqsecnum,floats,showpacs,nofootinbib,amsmath,amssymb,aps,prd,superscriptaddress]{revtex4}
\usepackage{amsmath}
\usepackage{amsfonts}
\usepackage{amssymb}
\usepackage[dvips]{graphicx}
\begin{document}

\newcommand\be{\begin{equation}}
\newcommand\ee{\end{equation}}
\newcommand\ba{\begin{eqnarray}}
\newcommand\ea{\end{eqnarray}}
\newcommand\bseq{\begin{subequations}} 
\newcommand\eseq{\end{subequations}}
\newcommand\bcas{\begin{cases}}
\newcommand\ecas{\end{cases}}
\newcommand{\p}{\partial}
\newcommand{\f}{\frac}
\newcommand{\nn}{\nonumber \\}
\def\rg{\rangle}
\def\lg{\langle}
\def\tr{{\rm Tr}}
\def\sgn{{\rm sgn}}
\def\intinf{\int_{-\infty}^{\infty}}
\def\lp{\ell_{\rm Pl}}
\def\CH{\hat{\mathcal{C}}_H}
\def\d{{\rm d}}
\def\La{\Lambda}
\def\Th{\Theta}
\def\de{\delta}
\def\ep{\epsilon}
\def\la{\lambda}
\def\vl{\vec{\la}}

\title{Local spinfoam expansion in loop quantum cosmology
\vskip10pt
}

\author{Adam Henderson}
\email{henderson@gravity.psu.edu}
\affiliation{\small Institute for Gravitation and the Cosmos \& Physics Department,
The Pennsylvania State University, University Park, PA 16802-6300, USA}
\author{Carlo Rovelli}
\email{rovelli@cpt.univ-mrs.fr}
     \affiliation{\small Centre de Physique Th\'eorique de Luminy\footnote{Unit\'e mixte de recherche (UMR 6207) du CNRS et des Universit\'es de Provence (Aix-Marseille I), de la M\'editerran\'ee (Aix-Marseille II) et du Sud (Toulon-Var); laboratoire affili\'e \`a la FRUMAM (FR 2291).}
     , Case 907, F-13288 Marseille, EU}
\author{Francesca Vidotto}
 \email{vidotto@cpt.univ-mrs.fr}
     \affiliation{\small Centre de Physique Th\'eorique de Luminy\footnote{Unit\'e mixte de recherche (UMR 6207) du CNRS et des Universit\'es de Provence (Aix-Marseille I), de la M\'editerran\'ee (Aix-Marseille II) et du Sud (Toulon-Var); laboratoire affili\'e \`a la FRUMAM (FR 2291).}
     , Case 907, F-13288 Marseille, EU}
     \affiliation{\small Dipartimento di Fisica Nucleare e Teorica,
        Universit\`a degli Studi di Pavia, and\\  Istituto Nazionale
        di Fisica Nucleare, Sezione di Pavia, via A. Bassi 6,
        I-27100 Pavia, EU}
\author{Edward Wilson-Ewing}
\email{wilsonewing@gravity.psu.edu}
\affiliation{\small Institute for Gravitation and the Cosmos \& Physics Department,
The Pennsylvania State University, University Park, PA 16802-6300, USA}

\begin{abstract}
\vskip10pt
The quantum dynamics of the flat Friedmann-Lema\^itre-Robertson-Walker
and Bianchi I models defined by loop quantum cosmology have recently been
translated into a spinfoam-like formalism.  The construction is facilitated
by the presence of a massless scalar field which is used as an internal clock.
The implicit integration over the matter variable leads to a nonlocal spinfoam
amplitude.  In this paper we consider a vacuum Bianchi I universe and show
that by choosing an appropriate regulator a spinfoam expansion can be obtained
without selecting a clock variable and that the resulting spinfoam amplitude is local. 

\end{abstract}

\pacs{04.60.Pp; 04.60.Kz, 98.80Qc}

\maketitle

\section{Introduction}
\label{s1}

The spinfoam formalism \cite{Perez:2003vx,Freidel:2007py,Engle:2007wy,Rovelli:2010wq} is a covariant approach to quantum gravity closely related to canonical loop quantum gravity (LQG) \cite{Rovelli,Thiemann,Ashtekar:1992tm,Rovelli:1987df, Rovelli:1989za}.  There are numerous results on the
relation between the two languages, for example in the context of the 3d theory \cite{Noui:2004iy},
of the full hamiltonian theory  \cite{Engle:2009ba,Han:2009ay,Han:2009aw,Dittrich:2009fb}, and in the context of cosmology \cite{Rovelli:2008dx,Battisti:2009kp, Bianchi:2010zs}, but more clarity is still needed. 

The relation between the canonical and the spinfoam languages can be studied in the simplified context of cosmology. The canonical quantum dynamics of cosmology is well under control in the loop quantization \cite{Ashtekar:2003hd,Ashtekar:2006wn,Ashtekar:2007em,Ashtekar:2009vc}, and a spinfoam expansion has been derived from the canonical theory for the flat Friedmann-Lema\^itre-Robertson-Walker (FLRW) \cite{Ashtekar:2009dn, Ashtekar:2010ve} and Bianchi type I \cite{Campiglia:2010jw} cosmologies. 
In both cases, a massless scalar field was included in the model. This matter field plays two related roles. First, it allows the theory to be straightforwardly deparametrized by treating it as an internal clock variable.  Second, it acts as a regulator since the implicit integration over it turns distributional transition amplitudes into regular functions. 
This procedure, however, leads to spinfoam amplitudes that are nonlocal in time.  A nonlocal spinfoam expansion can still be an effective computational 
tool, but does not match the structure of the spinfoam expression of the general theory  
\cite{Engle:2007uq,Livine:2007vk,Engle:2007qf,Freidel:2007py,Engle:2007wy}, where locality is a foundational principle and full covariance under the choice of clock-time variables is strictly implemented. 

In this paper we study how to obtain a path integral formulation for the loop quantum cosmology (LQC) of the vacuum Bianchi I space-time maintaining full covariance under the choice of clock-time variables, namely without deparametrizing the theory. We work without a matter field and introduce a regulator $\de$ to control the distributional aspect of the transition amplitudes. 
The physical inner product we define is accurate up to some small error which vanishes as $\de\to0$.
The idea of such a regulator was introduced in \cite{Rovelli:2009tp} for the case where the spectrum of the eigenvalues of the Hamiltonian constraint operator is discrete, but the regulator used  there is not appropriate in  the continuous case. In this paper we consider two different regulators which we call the Gaussian and the Feynman regulators. The choice of the regulator is not trivial: we show that only the second leads to \emph{local} amplitudes. 

The paper is organized as follows: in Section II we  briefly review the canonical
theory for the LQC of vacuum Bianchi I models \cite{Ashtekar:2009vc}. In Section III we derive two different regulated spinfoam expansions for that model and we close with a discussion of our results.

\section{The Loop Quantum Cosmology of the Vacuum \newline Bianchi Type I Model}
\label{s2}

We consider vacuum Bianchi I space-times, possibly with a nonzero cosmological constant $\La$.  
The metric for a Bianchi I space-time is 
\be \d s^2 = -N^2 \d t^2 + a_1(t)^2 \d x_1^2 + a_2(t)^2 \d x_2^2 + a_3(t)^2 \d x_3^2, \ee
where $N$ is the lapse and $a_i$ are the three directional scale factors.  In previous
treatments of the Bianchi I model in LQC, the lapse has been chosen to be $N = |a_1 a_2 a_3|$ and we make the same choice here.

We introduce a fiducial cell in order to deal with trivial divergences in the canonical analysis 
(the resulting physics is independent of the choice of the cell).  We use a cell adapted to the symmetries of the space-time: a rectangular cell with lengths $\ell_1, \ell_2$ and $\ell_3$ with respect to the fiducial metric $\d s_o^2 = \d x_1^2 + \d x_2^2 + \d x_3^2$.

The variables used in LQC are the Ashtekar connection $A_a^i$ and the densitized triad
$E^a_i$; in Bianchi I space-times they can be parametrized as follows \cite{Ashtekar:2009vc}:
\be A_a^i = \f{c^i}{\ell^i} (\d x^i)_a \qquad {\rm and} \qquad E^a_i = \f{p_i \, \ell_i}{V_o}
\left( \f{\p}{\p x_i} \right)^a, \ee
where there is no sum over the $i$ and $V_o = \ell_1\ell_2\ell_3$.
The Poisson brackets are given by
\be \{c_i, p_j\} = 8\pi G \gamma \delta_{ij}, \ee
and the $(c, p)$ variables are related to the ones in the metric by
\be p_1 = \sgn(a_1)|a_2a_3|\ell_2\ell_3 \qquad {\rm and} \qquad c_1 = \f{\gamma\ell_1}{NV_o}
\f{\d a_1}{\d t}, \ee
where $\gamma$ is the Barbero-Immirzi parameter.  The other relations can be obtained
via permutations.

Following the ideas of LQG, the fundamental variables of the kinematical Hilbert space
are holonomies of the connection and area fluxes of the densitized triads rather than
the connection and densitized triads themselves.  Accordingly, 
only operators corresponding to complex exponentials of $c_i$ are defined in the quantum theory,
no $\hat c_i$ operator exists.  Since the $p_i$ correspond to the physical areas of the
fiducial cell, they can be directly promoted to be operators.  See \cite{Ashtekar:2009vc} for
the details of the construction of the kinematical Hilbert space.  In the
$p_i$-representation the resulting Hilbert space is composed of functions
$\psi(p_1,p_2,p_3)$ with finite inner product
\be |\psi|^2 = \sum_{\vec{p} \: \in \: \mathbb{R}^3} \bar{\psi}(p_1,p_2,p_3) \psi(p_1,p_2,p_3). \ee
On these states $p_i$ act via multiplication and the holonomies $e^{i \mu_i c_i}$ act as
translations.  Finally, the kinematical inner product of basis states is
\be \lg p_1, p_2, p_3 | p_1^\prime, p_2^\prime, p_3^\prime \rg = \delta_{p_1p_1^\prime}
\delta_{p_2p_2^\prime}\delta_{p_3p_3^\prime}, \ee
note that the $\delta$'s are Kronecker delta functions, not Dirac delta distributions.

The Hamiltonian constraint operator has been derived in \cite{Ashtekar:2009vc}. Before writing the explicit form of the operator itself, we make a change of coordinates which simplifies its form. We introduce the $\la_i$ variables defined as
\be p_i = \sgn(\la_i) (4\pi\gamma\sqrt\Delta\lp^3)^{2/3} \la_i^2, \ee
where $\Delta\lp^2 = 4\sqrt{3}\pi\gamma \lp^2$ is the area gap in LQC, 
the minimum eigenvalue of the area operator.  Next, let 
\be v = 2 \la_1\la_2\la_3, \ee
which is proportional to the physical volume of the fiducial cell.

Any state%
\footnote{This is not true for states corresponding to singular geometries, but since these states
decouple under the dynamics their behavior is trivial and not important for this work.}
can now be described by a ket $|\la_1, \la_2, v \rg$. The kinematical inner product is 
\be \lg \la_1, \la_2, v | \la_1^\prime, \la_2^\prime, v^\prime \rg =
\delta_{\la_1\la_1^\prime} \delta_{\la_2\la_2^\prime}\delta_{vv^\prime}. \ee

As shown in \cite{Ashtekar:2009vc}, the Hamiltonian constraint operator for the vacuum Bianchi I model (with a minor
modification in order to include the cosmological constant $\La$) is%
\footnote{Although in general the variables $\la_i$ and $v$ can be negative as well as positive,
it is possible to only consider the positive case due to the parity properties of the wave function,
see \cite{Ashtekar:2009vc} for details.  This is what is done here.}
\begin{align} \CH \Psi(\la_1, \la_2, v) =& \f{\pi\hbar\lp^2}{16} \Bigg[
(v+2) \sqrt{v(v+4)}\, \Psi^+_4 (\la_1,\la_2,v) - v(v+2)\, \Psi^+_0 (\la_1,\la_2,v) \nn &
-v(v-2)\, \Psi^-_0 (\la_1,\la_2,v) +(v-2) \sqrt{v|v-4|}\, \Psi^-_4 (\la_1,\la_2,v) \nn &
+ 4\gamma^2\Delta\La\lp^2 v^2 \: \Psi(\la_1,\la_2,v) \Bigg], \end{align}
where the $\Psi^\pm_{0,4}$ are defined as follows:
\begin{align} \Psi^\pm_n(\la_1,\la_2,v)=& \:\Psi\left(\f{v\pm n}{v\pm2}\cdot
\la_1,\f{v\pm2}{v}\cdot\la_2,v\pm n\right)+\Psi\left(\f{v\pm n}{v\pm2}\cdot\la_1,
\la_2,v\pm n\right)\nn & +\Psi\left(\f{v\pm2}{v}\cdot\la_1,\f{v\pm n}{v
\pm2}\cdot\la_2,v\pm n\right)+\Psi\left(\f{v\pm2}{v}\cdot\la_1, \la_2,v\pm n
\right) \nn & +\Psi\left(\la_1,\f{v\pm2}{v}\cdot\la_2,v\pm n\right)+
\Psi\left(\la_1,\f{v\pm n}{v\pm2}\cdot\la_2,v\pm n\right).
\label{hamiltoniana}
 \end{align}

The Hamiltonian constraint defines the physical inner product.  There are two ways that have been used to construct the physical inner product in cosmology. The first is to ``deparametrize" the theory by choosing one of the variables as a time-variable, and expressing the dynamics of the other variables with respect to it.  This is commonly done in loop quantum cosmology by coupling a massless scalar field to the metric and using it as a clock. 

The second possibility, which is the one we use here, is to maintain explicit covariance under the choice of the independent variable and define the physical physical inner product by ``group averaging".   Heuristically, given a suitable state in the kinematical
Hilbert space $| \phi \rg$, a physical state is obtained by
acting on the states with a delta function of the constraint,
$ |\Phi \rg_{\rm phy} = \delta(\CH) | \phi \rg$.  	More precisely, 
the physical inner product  can be defined by the expression
\be \lg \Psi | \Phi \rg_{\rm phy} = \f{1}{2\pi} \intinf \d\alpha \: \lg \psi | e^{i\alpha\CH} |
\phi \rg_{\rm kin}. \ee

Denoting $(\la_1, \la_2, v)$ by $\vl$, the physical inner product between the physical
states generated from the kinematic basis states for the vacuum Bianchi I model in LQC is given by
\be \label{phys-ip} \lg \vl_F | \vl_0 \rg_{\rm phy} =
\f{1}{2\pi} \intinf \d\alpha \:
\lg \vl_F | e^{i\alpha \CH} | \vl_0 \rg_{\rm kin}. \ee
This expression may contain divergences. In the next section we reformulate it so that it can be computed explicitly.

\section{The Vertex Expansion}
\label{s3}

In this section we express the physical inner product via a Feynman path integral
construction with the form of a vertex expansion, much like spinfoam models.
In the first part, we   review the procedure presented in \cite{Ashtekar:2009dn, Ashtekar:2010ve,Campiglia:2010jw} for the FLRW and Bianchi I models and show how, when it is na\"ively followed,
the final result fails to be well defined for vacuum space-times.  In the following
subsections, we present two examples of how one can introduce an external regulator
as proposed in \cite{ Rovelli:2009tp} in order to obtain a well-defined result.

\subsection{The Standard Procedure}
\label{s3.1}

As usual in a Feynman path integral construction, we break up the action of $\CH$ into
$N$ equal parts (i.e., $\exp(i\alpha\CH) = [\exp(i\alpha\CH/N)]^N$) and insert a completeness
relation between each term:
\begin{align} \lg \vl_F | \vl_0 \rg_{\rm phy} &= \f{1}{2\pi} \intinf \d \alpha \lg \vl_N | e^{i\alpha \CH}
| \vl_0 \rg \nn
&= \f{1}{2\pi} \intinf \d\alpha \sum_{\vl_1}\sum_{\vl_2}\ldots\sum_{\vl_{N-1}}
\lg \vl_F | e^{i\ep\CH} | \vl_{N-1} \rg \ldots
\lg \vl_2 | e^{i\ep\CH} | \vl_1 \rg
\lg \vl_1 | e^{i\ep\CH} | \vl_0 \rg,
\end{align}
where we have introduced $\ep = \alpha/N$.  Since this relation holds for any $N$, it is
straightforward to take the limit of $N\to\infty$. However, an important point here is that
the $N\to\infty$ limit is \emph{inside} the integral over $\alpha$.

Now, in the limit of small $\ep$ (or large $N$) one immediately sees that (up to higher order
terms in $\ep$ which can safely be neglected)
\be \lg \vl | e^{i\ep\CH} | \vl \rg \approx e^{i\ep\Th_{\vl\vl}} \qquad {\rm and} \qquad
\lg \vl | e^{i\ep\CH} | \vl^\prime \rg \approx i\ep\Th_{\vl\vl^\prime}, \ee
where $\Th_{\vl\vl^\prime} = \lg \vl |\CH| \vl^\prime \rg$ are the matrix elements of $\CH$
and in the second relation we have assumed that $\vl \ne \vl^\prime$.

We   now reformulate the inner product in terms of a vertex expansion.  This
is done by rewriting the sum in terms of the number $M$ of transitions in $\vl$, that is to say
the number of times $\vl_{i+1} \ne \vl_i$.  Denoting the `time'-step of the $i$-th transition
by $N_i$, one finds that (see \cite{Ashtekar:2009dn, Ashtekar:2010ve} for a derivation of this result)
\begin{align} \lg \vl_F | \vl_0 \rg_{\rm phy} = \f{1}{2\pi} \intinf \d\alpha
\lim_{N\to\infty} & \sum_{M=0}^N \sum_{N_i} \sum_{\vl_{M-1}}\ldots\sum_{\vl_1} (i\ep\
Th_{\vl_F\vl_{M-1}}) \ldots (i\ep\Th_{\vl_1\vl_0}) \nn & (e^{i\ep\Th_{\vl_F\vl_F}})^{N-N_M-1}
(e^{i\ep\Th_{\vl_{M-1}\vl_{M-1}}})^{N_M-N_{M-1}-1}\ldots \nn &
(e^{i\ep\Th_{\vl_1\vl_1}})^{N_2-N_1-1}
(e^{i\ep\Th_{\vl_0\vl_0}})^{N_1-1},
\end{align}
where we have introduced the shorthand
\be \sum_{N_i} = \sum_{N_M=M}^{N-1} \: \sum_{N_{M-1}=M-1}^{N_M-1} \ldots \sum_{N_2=2}^{N_3-1}
\sum_{N_1=1}^{N_2-1}. \ee
In the $N\to\infty$ limit, the sums become integrals and the vertex expansion for the physical
inner product is given by
\be \label{intalpha} \lg \vl_F | \vl_0 \rg_{\rm phy} = \f{1}{2\pi} \intinf \d\alpha
\sum_{M=0}^\infty \sum_{\vl_{M-1}}\ldots\sum_{\vl_1} \Th_{\vl_F\vl_{M-1}} \ldots
\Th_{\vl_1\vl_0} A(\vl_1, \ldots, \vl_{M-1}; \alpha), \ee
where
\begin{align} \label{amp1} A(\vl_1, \ldots, \vl_{M-1}; \alpha) =& i^M \int_0^\alpha \d t_M \int_0^{t_M}
\d t_{M-1} \ldots \int_0^{t_2} \d t_1 (e^{i\Th_{\vl_F\vl_F}})^{\alpha-t_M} \nn & \qquad
(e^{i\Th_{\vl_{M-1}\vl_{M-1}}})^{t_M - t_{M-1}} \ldots
(e^{i\Th_{\vl_1\vl_1}})^{t_2-t_1} (e^{i\Th_{\vl_0\vl_0}})^{t_1}. \end{align}
The integrals in this expression can be evaluated, the general solution
is (see \cite{Ashtekar:2010ve} for details)
\be \label{ampG} A(\vl_1, \ldots, \vl_{M-1}; \alpha)\, =
\Bigg[ \prod_{k=1}^p \frac{1}{(n_k-1)!} \left( \frac{\partial}{\partial
\Theta_k} \right)^{n_k-1}\, \Bigg] \, \sum_{i=1}^p
\frac{e^{i \alpha \Theta_i}}{\prod_{j \neq i}^p
( \Theta_i -  \Theta_j)}, \ee
where $\Theta_i$ are the $p$ distinct values of $\Th_{\vl_i\vl_i}$ taken along the history and $n_i$ is
the number of times the value $\Theta_i$ is repeated.  In the simplest case of all the $n_i$ being 1,
this gives%
\footnote{Even though expression \eqref{ampG} is well defined in the limit that some of the diagonal matrix elements of the constraint are zero, the integral over $\alpha$ in Eq.~\eqref{intalpha} is divergent instead of distributional in this limit.  For this reason, we have added a cosmological constant to the model since the constraint has no diagonal terms if $\Lambda = 0$.  Alternatively, we could have instead considered a different basis in the kinematical Hilbert space or worked with the master constraint.} 
\be A(\vl_1, \ldots, \vl_{M-1}; \alpha) = \sum_{i=0}^M \f{e^{i\alpha\Th_i}}
{\prod_{j\ne i} (\Th_i - \Th_j)}~. \ee

The last step is to perform the integral over $\alpha$ and
express $\lg \vl_F | \vl_0 \rg_{\rm phy}$ as a sum over $M$ of some function of the intermediate
steps $\vl_i$.  However, this requires pulling the infinite sum over $M$ outside of the integral
and this causes major difficulties: in the simplest case where all $\vl_i$ are
different, one obtains a sum of Dirac delta distributions, \emph{not a function}.  In the more
general cases, the situation is even worse as one obtains derivatives of Dirac delta distributions
as well.  While this provides a formal perturbative solution to the constraint, it is distributional
term by term.  This would give a sum over distributions outside of any integral and the vertex
expansion presented here would fail to be well defined.  This procedure is carried out in the
Appendix for the sake of completeness.

If a massless scalar field is added then the distributions appear under an integral over the scalar
field momentum.  Similarly, any degree of freedom with a continuous spectrum introduces integrals
which   give meaning to the distributions which arise.  It is the totally discrete nature of the 
kinematic Hilbert space in the vacuum case that gives rise to the ill-defined expansion above.
As there is no matter in this case, we introduce an external regulator $\de$ in order to obtain
functions rather than distributions in the expression of the physical inner product.  
There are two
natural regularizations which we   present here:  a
Gaussian suppression of the integral at infinity   and
an exponential suppression \`a la Feynman.

\subsection{Gaussian Regulator}
\label{s3.2}     

Recall that the group averaging procedure can be thought of as putting a Dirac delta
distribution projector in the kinematical inner product.  A natural way to regulate
the vertex expansion is to approximate the Dirac delta distribution by a Gaussian with a
small spread $\de$.
The approximated physical inner product given by this regulated group averaging is
\begin{align} \lg \vl_F | \vl_0 \rg_{\rm G,\de} & = \f{1}{2\pi} \intinf \d\alpha \:
\lg \vl_F | e^{i\alpha \CH - \de^2 \alpha^2} | \vl_0 \rg_{\rm kin}. \end{align}
While the projector given by group averaging kills all of the parts of the given wave
function that don't satisfy the constraint, this regulated `projector' exponentially
damps all of the parts of the wave function that do not satisfy the constraint.  Clearly,
the correct projector is recovered in the limit $\de\to0$.

Following the same procedure outlined above in order to obtain a vertex
expansion for this regulated inner product, we obtain
\be \lg \vl_F | \vl_0 \rg_{\rm G,\de} = \f{1}{2\pi} \intinf \d\alpha \sum_{M=0}^\infty
\sum_{\vl_i} \Th_{\vl_F\vl_{M-1}} \ldots \Th_{\vl_1\vl_0}
e^{-\de^2 \alpha^2} A(\vl_1, \ldots, \vl_{M-1}; \alpha), \ee
where $A(\vl_1, \ldots, \vl_{M-1}; \alpha)$ is again given by Eq.~(\ref{ampG}).
Pulling the sum over $M$ outside of the integral and 
evaluating the integral over $\alpha$, we have
\be \lg \vl_F | \vl_0 \rg_{\rm G,\de} = \f{1}{2\pi}  \sum_{M=0}^\infty
\sum_{\vl_i} \Th_{\vl_F\vl_{M-1}} \ldots \Th_{\vl_1\vl_0}
\bar{A}_{\rm G,\de} (\vl_1, \ldots, \vl_{M-1}), \ee
note that $\bar{A}(\vl_1, \ldots, \vl_{M-1})$ represents the
``group-averaged'' $A(\vl_1, \ldots, \vl_{M-1}; \alpha)$.  For the
Gaussian regulator presented here, one can see that
\be \bar{A}_{\rm G,\de} (\vl_1, \ldots, \vl_{M-1}; \alpha)\, =
\prod_{k=1}^p \frac{1}{(n_k-1)!} \left( \frac{\partial}{\partial
\Theta_k} \right)^{n_k-1}\,\, \sum_{i=1}^p
\frac{\sqrt{\pi}}{\delta} \frac{e^{- \Theta_i^2/ 4 \delta^2}}{\prod_{j \neq i}^p
( \Theta_i -  \Theta_j)}.
\ee
Again the $\Th_i$ label the $p$ distinct values of  $\Th_{\vl \vl}$ taken along the history
($\vl_0, \ldots, \vl_M$) and $n_i$ the number of times that value is repeated.  Notice that
one cannot take the limit of $\de\to0$ at this point, the fact that $\de$ is nonzero is
absolutely necessary in order to pull the sum over $M$ outside of the integral.

The integral here gives Gaussians and derivatives thereof instead of Dirac delta
distributions and derivatives thereof.  The regulated physical inner product is then well
defined and is a good approximation:
\be \lg \vl_F | \vl_0 \rg_{\rm phy} \approx \lg \vl_F | \vl_0 \rg_{\rm G,\de}. \ee
This expression can be used as a computational tool but since it is nonlocal, it
cannot be viewed as a building block for the local spinfoam expansion of full quantum
gravity.

\subsection{Feynman Regulator}
\label{s3.3}

Another possible approach is to split the integral given in Eq.~(\ref{phys-ip}) into
integrals over positive and negative
$\alpha$ and regulate each integral separately, much as is done for the Feynman propagator,
with an exponential suppression as follows:
\ba \lg \vl_F | \vl_0 \rg_{\rm F,\de} &=&
\f{1}{2\pi} \int_0^\infty \d\alpha \: \lg \vl_F | e^{i\alpha \CH - \delta \alpha} | \vl_0 \rg_{\rm kin}
+ \f{1}{2\pi} \int_{-\infty}^0 \d\alpha \: \lg \vl_F | e^{i\alpha \CH + \delta \alpha} | \vl_0 \rg_{\rm kin} \\
&=& \lg \vl_F | \vl_0 \rg_{\rm +,\delta}+ \lg \vl_F | \vl_0 \rg_{\rm -,\delta}, \ea
this approximates $\lg \vl_F | \vl_0 \rg_{\rm phy}$ up to some small error which vanishes as
$\de\to0$.

As an aside, it is interesting to note that if we construct a phase space path integral for
this model as done for FLRW in \cite{achwkb} we find that
each half is related to fixing one sign for the lapse or fixing a single direction for the
time evolution.  It is necessary though to include both terms to have a solution to the constraint.  

Now, the integral over negative $\alpha$ is the complex conjugate of that over positive $\alpha$,
so for simplicity we first focus on just one half of this expression and see what effect
the regularization has on the integrals over $\alpha$.  Taking
\be \lg \vl_F | \vl_0 \rg_{\rm +,\de} = \f{1}{2\pi} \int_0^\infty \d\alpha \:
\lg \vl_F | e^{i\alpha \CH - \de \alpha} | \vl_0 \rg_{\rm kin}, \ee
the expansion of $\lg \vl_F | e^{i\alpha \CH} | \vl_0 \rg_{\rm kin}$ can be carried out as before
and then the integral over $\alpha$ gives
\be \lg \vl_F | \vl_0 \rg_{+,\de} = \f{1}{2\pi} \sum_{M=0}^\infty \sum_{\vl_{M-1}}\ldots\sum_{\vl_1}
\Th_{\vl_F\vl_{M-1}} \ldots \Th_{\vl_1\vl_0} \bar{A}_{+,\de}(\vl_1, \ldots, \vl_{M-1}), \ee
where the amplitude is given by
\be \bar{A}_{+,\de} (\vl_1, \ldots, \vl_{M-1}) = \int_0^\infty \d\alpha\; e^{-\delta \alpha}
A(\vl_1, \ldots, \vl_{M-1}; \alpha), \ee
and $A(\vl_1, \ldots, \vl_{M-1}; \alpha)$ is given by Eq.~(\ref{ampG}).

Surprisingly, the integral over alpha reduces the very nonlocal expression given in Eq.~(\ref{ampG})
with a simple \emph{local} one being just a product of matrix elements:
\be \bar{A}_{+,\de}(\vl_1, \ldots, \vl_{M-1}) = \frac{i (-1)^M} {\prod_{m=0}^M (\Th_{\vl_m \vl_m}+ i \delta)}. \ee
Putting all of the pieces together, the expansion for the positive half is
\be \lg \vl_F | \vl_0 \rg_{+\delta} = \f{1}{2\pi} \sum_{M=0}^\infty \sum_{\vl_{M-1}}\ldots\sum_{\vl_1}
\frac{i (-1)^M \Th_{\vl_F\vl_{M-1}} \ldots \Th_{\vl_1\vl_0}}{(\Th_{\vl_M \vl_M}+ i \delta) \ldots
(\Th_{\vl_0 \vl_0}+ i \delta)}. \label{vivazapata}\ee
%
The $\delta$ remains non-zero in the resulting expansion, since in the limit that $\delta$ goes to zero it reduces to a sum of distributions.
The physical inner product, regulated by the parameter $\delta$, is given by the sum of this expansion
with its complex conjugate.  Equivalently,
\be \lg \vl_F | \vl_0 \rg_{\rm phy} = 2 \, \Re \Big( \lg \vl_F | \vl_0 \rg_{+\delta} \Big) + {\rm err}(\de), \ee
where ${\rm err}(\de)$ is a small error term which vanishes in the limit of $\de\to0$.
Notice that \eqref{vivazapata} can be rewritten in the form
\be \lg \vl_F | \vl_0 \rg_{+\delta} = \f{i}{2\pi} ~ \sum_{M=0}^\infty \ \ 
\sum_{\vl_{1}...\vl_{M\!-\!1}}
\ \ \prod_{f}  A_f(\vl_{f})\ \ 
\prod_{v} A_v(\vl_{f}), \label{vivaelche}
\ee
where $f=0,...,M$;  $v=1,...,M$; the ``face" amplitude is $A_f= (\Th_{\vl_f \vl_f}+ i \delta)^{-1}$, the ``vertex" amplitude is $A_v=-\Th_{_{\vl_v \vl_{v-1}}}$, and we write $\vl_M=:\vl_F$. The expression \eqref{vivaelche} is precisely the general expression for (local) spinfoams, if we identify the sum over spinfoam two-complexes with the sum over $M$, the sum over colorings with the sum over the $\vl_f$, the spinfoam vertices with the transitions, and the spinfoam faces with the sequences of steps without transitions.

\vskip8pt


The fact that
the Feynman regularization gives an amplitude which is
local is a remarkable result. In the previous works 
\cite{Ashtekar:2009dn, Ashtekar:2010ve,Rovelli:2009tp,Campiglia:2010jw},
the reconstruction of the spinfoam-like expansion obtained
from LQC lacks locality in the resulting expression;
here locality means that the amplitude of one history is the
product of amplitudes of the elements forming the history.
Notice that in quantum mechanics locality is always present
in sums over histories.  For instance, in quantum electrodynamics Feynman
graphs, the amplitude of a graph is the product of the
amplitudes of the vertices and of the propagators.  In the functional
integral formulation of quantum mechanics, the amplitude is the exponent
of an integral, which is to say (morally) the product of
exponents: again a product of local terms.  
At least in the Feynman graph integral, and possibly in other contexts, a nonlocal measure might indicate that something is wrong.
Thus, this type of locality is a desired feature of a sum over
histories formulation of quantum gravity.

A na\"ive  regularization is not enough to obtain locality,
as we have seen in the case of the Gaussian regulator where
one again obtains a nonlocal amplitude.
With the Feynman regularization, the amplitude
is local.

\section{Discussion}
\label{s4}

We have shown that in loop quantum cosmology it is possible to introduce a regulator in order to approximate the vertex expansion of the physical inner product and that one such regulator gives a local expression.
The strategy to calculate the physical inner product is to choose a small, nonzero
$\de$ and to perform one of the regulated expansions presented here
to some number of transitions $m$ and this approximates the true physical inner
product to some order determined by the regulator $\de$ and the number of
transitions $m$.  A priori, the convergence properties of the series is unknown.

The idea of introducing an external
regulator was proposed in \cite{Rovelli:2009tp} in the context where the Hamiltonian constraint operator has a discrete
spectrum.  In that case, the regulator removes actual divergences in the expansion. In the case
of a continuous spectrum, the terms in the expansion are simply distributional.
The regularizations introduced here are natural from the point of view of handling 
distributions.

The Feynman regularization adds an additional similarity to spin foams.  The spin
foam amplitude for a single triangulation and set of labels, for a single history of spin networks,
is given by a product of amplitudes associated to each vertex, face, and edge.  
In contrast, for the Gaussian regulator presented here and for the vertex expansion of FLRW with a massless scalar field given in \cite{Ashtekar:2009dn, Ashtekar:2010ve}  the amplitude for a single discrete history is not a simple product of amplitudes, rather it is a nonlocal expression depending on the properties of the entire history.  By not integrating away the matter degrees of freedom from the bulk, and using the Feynman regulator introduced here, the amplitude is reduced to a simple local product.

The local form of the amplitude makes the relation between the expansion of the LQC scalar products and the spinfoam formalism transparent:  the sum over the number $M$ of transitions is recognized
as the analog of the sum over two-complexes, the sum over the $\vl_f$ as the sum over spinfoam colorings, the transitions as the spinfoam vertices and the sequences of steps without transitions as the spinfoam faces.

In addition, we observe that by using the Feynman regulator in the models studied in
\cite{Ashtekar:2009dn, Ashtekar:2010ve, Campiglia:2010jw} \emph{before} integrating over the
momentum of the scalar field, one obtains an integrand which is well-defined on its own
(rather than distributional) and which gives a local expression for the amplitude, thus
providing a local extension to the results of these previous works.
\vskip2,5cm

\section*{Acknowledgments}

This work was supported by the NSF grant PHY0854743, the George A.\ and Margaret
M.\ Downsbrough Endowment, the Eberly research funds of Penn State, Le Fonds
qu\'eb\'ecois de la recherche sur la nature et les technologies, and the Edward
A.\ and Rosemary A.\ Mebus funds.  E.\ W.-E.\ thanks the CPT for their hospitality
during his visit there.
\vskip2cm

\begin{appendix}

\section*{APPENDIX: \ Distributional Expansion}

If we ignore the problems mentioned in Sec.~\ref{s3.1} and evaluate the integral over $\alpha$ without
introducing any sort of regulator, the resulting expansion is
\be \lg \vl_F | \vl_0 \rg_{\rm phys} =  \sum_{M=0}^\infty \sum_{\vl_{M-1}}\ldots\sum_{\vl_1}
\Th_{\vl_F\vl_{M-1}} \ldots \Th_{\vl_1\vl_0} \bar{A}(\vl_1, \ldots, \vl_{M-1}), \ee
where
\be  \bar{A}(\vl_1, \ldots, \vl_{M-1})  = \f{1}{2\pi} \intinf \d\alpha \; A(\vl_1, \ldots, \vl_{M-1};
\alpha). \ee
This integral can be easily evaluated giving
\be \bar{A}(\vl_1, \ldots, \vl_{M-1})  = \Bigg[ \prod_{k=1}^p \frac{1}{(n_k-1)!} \left( \frac{\partial}
{\partial \Theta_k} \right)^{n_k-1} \Bigg]\, \sum_{i=1}^p \frac{\delta(\Theta_i)}{\prod_{j \neq i}^p
( \Theta_i -  \Theta_j)}, \ee
where, as before, $\Theta_i$ are the $p$ distinct values of $\Th_{\vl_i\vl_i}$ taken along the history and
$n_i$ is the number of times the value $\Theta_i$ is repeated.  One can verify that this distribution is
equivalent to the following simpler one:
\be \bar{A}(\vl_1, \ldots, \vl_{M-1})  = (-1)^{M+1} \,\, \sum_{i=1}^p \frac{(-1)^{n_i}}{(n_i-1)!} \left(
\frac{\partial}{\partial \Theta_i} \right)^{n_i-1} \frac{\delta(\Theta_i)}{\prod_{j \neq i}^p
\Theta_j^{n_j}}. \ee
At a formal level, this sum of distributions provides a perturbative solution to the constraint.
\vskip10mm
\end{appendix}


\end{document}